\documentclass[preprint2]{aastex}
\usepackage{natbib}
\usepackage[dvips]{color}
\newif\ifAMStwofonts
\AMStwofontstrue

\def\vB{{\bf B}}

\def\gsim{~\rlap{$>$}{\lower 1.0ex\hbox{$\sim$}}}

\def\simpropto{\lower.2ex\hbox{$\; \buildrel \propto \over \sim \;$}}
\def\ltsim{\lower.5ex\hbox{$\; \buildrel < \over \sim \;$}}
\def\gtsim{\lower.5ex\hbox{$\; \buildrel > \over \sim \;$}}
\def\ltsim{\lower.5ex\hbox{$\; \buildrel < \over \sim \;$}}
\def\gtsim{\lower.5ex\hbox{$\; \buildrel > \over \sim \;$}}

\def\kms{\mbox{km\,s$^{-1}$}}

\def\dd{\,{\rm d}}

\def\kms{\ {\rm km\,s^{-1}}}
\def\hmpc{\ {\rm h^{-1}Mpc}}

\def\dd{{\rm d}}

\def\ln{{\rm ln}}

\def\pmb#1{\setbox0=\hbox{#1}%
\kern-.025em\copy0\kern-\wd0
\kern.05em\copy0\kern-\wd0
\kern-.025em\raise.0433em\box0}

\def\vr{\pmb{$r$}}

\def\simlt{\lower.5ex\hbox{$\; \buildrel < \over \sim \;$}}
\def\simgt{\lower.5ex\hbox{$\; \buildrel > \over \sim \;$}}

\newcommand{\beq}{\begin{equation}}
\newcommand{\eeq}{\end{equation}}
\def\beqa{\begin{eqnarray}}
\def\eeqa{\end{eqnarray}}
\def\fixit#1{}
\def\hmpc{h^{-1}\,{\rm Mpc}}

\def\dd{{\rm d}}

\def\mudm{\mu_{_{\delta\! M_*}}}
\def\sigdm{\sigma_{_{\delta\! M_*}}}
\def\mual{\mu_{_{\delta\! \alpha}}}
\def\sigal{\sigma_{_{\delta\! \alpha}}}

\shorttitle{Photometric bulk flows }
\shortauthors{Nusser, Branchini \& Davis}
\begin{document}
\title{Bulk flows from galaxy luminosities: {\large application to 2MASS redshift survey and
forecast for next-generation datasets}}
\author{Adi Nusser\altaffilmark{1}}
\affil{Physics Department and the Asher Space Science Institute-Technion, Haifa 32000, Israel}
\author{Enzo Branchini\altaffilmark{2}}
\affil{Department of Physics, Universit\`a Roma Tre, Via della Vasca Navale 84, 00146, Rome, Italy}
\author{Marc Davis\altaffilmark{3}}
\affil{Departments of Astronomy \& Physics, University of California, Berkeley, CA. 94720}
\altaffiltext{1}{E-mail: adi@physics.technion.ac.il}
\altaffiltext{2}{E-mail: branchin@fis.uniroma3.it}
\altaffiltext{3}{E-mail: mdavis@berkeley.edu}


\begin{abstract}

We present a simple method for measuring cosmological bulk flows from 
large redshift surveys, based on the apparent dimming or brightening of galaxies
due to their peculiar motion. 
It is  aimed at estimating bulk flows of  cosmological volumes
 containing  large numbers of galaxies. Constraints on the bulk flow 
 are obtained by minimizing systematic variations in galaxy luminosities
with respect to a reference luminosity function measured 
from the whole survey.  
This method offers two advantages over more popular bulk flow estimators:
it is independent of error-prone distance indicators and of the poorly-known galaxy bias.
We apply the method to the
 2MASS redshift survey (2MRS) to
measure the local bulk flows of spherical shells centered on the Milky Way (MW). 
The result is consistent with that obtained by Nusser and Davis (2011)
using the SFI++ catalogue of Tully-Fisher distance indicators.
 We also make an assessment of the ability of the method to constrain   bulk flows at larger redshifts ($z=0.1-0.5$)
from next generation datasets. As a case study we consider the 
planned EUCLID survey. Using this method we  will be able to measure 
a bulk motion of $ \sim 200 \kms$ of $10^6$ galaxies with photometric redshifts,
at the $3\sigma$ level for both  $z\sim 0.15$ and $z\sim 0.5$. 
Thus the method will 
allow us to   put strong constraints on 
dark energy models as well as alternative theories for structure formation. 

 \end{abstract}

\keywords{Cosmology}

\section{Introduction}

In the standard cosmological paradigm, deviations from the Hubble flow, 
commonly  named
peculiar motions, are the result of the gravitational force field dominated by
 the dark matter, 
with luminous galaxies behaving like test particles.
Hence observations of the peculiar velocity field of galaxies are a direct probe of the three
 dimensional distribution of the dark matter. 
Cosmological bulk flows of spherical volumes  centered on the MW
are the most intuitive moments of this velocity field. 
These bulk flows are typically measured over
large spheres (i.e. $\gtsim 30\hmpc$) where {\em linear} 
 gravitational instability holds \citep[e.g.][]{n91}, facilitating   comparisons with cosmological 
  models. 
%

So far,  estimates  of the bulk flow in the local Universe within $\sim 100\hmpc$ from us have mostly  relied  on galaxy peculiar velocities  inferred from known distance indicators and measured redshifts. 
The distance indicators commonly used to estimate peculiar velocities 
 are based on  well defined, if often heuristic, 
relations between intrinsic, observable galaxy properties, one of which depends on the galaxy distance.
The typical example is the Tully-Fisher relation \citep{TF77} between the absolute magnitudes and the rotational velocity of spiral galaxies. 
Distance indicators have been extremely useful in enforcing our 
confidence in gravitational instability as the main mechanism for structure formation \citep{DN10} and in putting constraints on cosmological models. 
A recent analysis by \cite{ND11} has shown that the  bulk flow inferred from a trimmed version of 
 the SFI++  catalog of spiral
galaxies with I-band Tully-Fisher distances \citep{mas06, spring07, DN10} is  consistent with the standard $\Lambda$CDM cosmological model with  the best fit parameters of the  Seven-Year Wilkinson Microwave Anisotropy Probe  (WMAP7)   \cite[e.g.][]{jaro10,wmap7}. Furthermore,
using the recovered bulk flow, \cite{ND11} derived
an interesting constraint on the growth rate of fluctuations in the Universe at $z\sim0$. 

The paucity of distance indicators available and the observational difficulty in
measuring the relevant quantities make peculiar velocity quite difficult to estimate.
In addition, their accuracy degrades linearly with redshift, limiting their usefulness to rather 
small redshifts ($cz<10000\kms$). 
More accurate peculiar velocity measurements 
based on Ty1a SN \citep[e.g.][]{r97,Dai11} and surface brightness  \citep[e.g.][]{SBr99} are possible  for substantially fewer galaxies. 
As a result, galaxy peculiar velocities {  have been estimated} for a relatively small number of galaxies.
To make the matter worse these galaxies are not uniformly distributed across the sky, 
potentially inducing unwanted {  biases} in the bulk flow estimate.
Here we propose to circumvent these {  potential} problems {  induced by the use
of distance indicators}  and constrain the bulk flow {  using only galaxy luminosities and redshifts}.

Redshifts of galaxies systematically differ from the actual   distances
by the line-of-sight component of the bulk motion. {  Hence, the intrinsic luminosities of galaxies 
inferred from the observed flux using redshifts rather than distances 
redshifts appears to be brighter or dimmer. In presence of large bulk flows the 
effect is systematic and its strength depends on
on amplitude and  direction  of the bulk motion.}
Object-by-object magnitude variations as a result of peculiar motions are overwhelmed by
the natural spread in the distribution of magnitudes.  Hence, individual peculiar velocities 
cannot be derived from these considerations. However, an estimate of the  bulk motion of 
some sobvolume within the survey can be derived on a statistical basis, by comparing the 
 luminosity distribution of galaxies in the subvolume with the one
 in the whole survey.

The method has a long history.   \cite{TYS} correlated the  magnitudes with redshifts 
of galaxies to constrain the velocity of the Virgo cluster relative to the Local Group (LG) of 
galaxies. The main limitation of this method has been the limited number of the objects available 
and the limited size of the sampled volume.
The extension we propose here is timely in view of the considerable progress 
in {  current and future} redshift surveys. 

In \S2  we introduce the method, presenting  general expressions and deriving the 
relevant approximations. In \S3 we apply the method to the Two Mass redshift Survey (2MRS) 
of $\sim 23000$ galaxies limited to magnitude $K=11.25$.
In \S4 we discuss  prospects for successful applications of the method to future redshift surveys.  We conclude in  \S5 with a general discussion.

\section{The method}
\subsection{The  set-up}

Consider a subvolume in a 
survey of galaxies with measured redshifts $cz$ (in $\kms$) and apparent magnitudes 
$m$  limited to  $m<m_l$. The  redshifts are given in the frame of reference of 
the cosmic microwave background (CMB) radiation. 
We envisage two descriptions for the geometry of the subvolume. 
{\it (i)} a local subvolume of a thick shell centered on the observer (the MW) in an all sky redshift survey, and{\it (ii)} a  distant region where all galaxies  closely lie along the same line-of-sight. 

Let $r$ (also in $\kms$) be the  (unobserved) 
luminosity distance to a galaxy. For simplicity of notation and description we assume here that the distance and spatial 
extent of the survey are small so that $r$ is well approximate by the physical distance. 
The results can  readily be extended to the general case once we specify the underlying 
cosmological model.
The unknown, {\it true} absolute magnitude, $M$,  of a galaxy is expressed as
\begin{equation}
M=m-15 -5\log r =  M_0 - \gamma
\label{eq:shift}
\end{equation} 
where the {\it measurable} absolute
 magnitude $M_0=m-15-5 \log cz$ is determined from observations, and $\gamma \equiv 5\log (r/cz)$. The observed redshift of the object is $cz=r+v$ where $v$ is the line-of-sight component of its
 peculiar velocity.
A bulk motion of the subvolume
yields a systematic difference between $r$ and $cz$ which induces a mismatch between 
$M$ and $M_0$.
Therefore, the  bulk motion of the subvolume (relative to the motion of the whole survey) can be 
estimated by demanding that 
the distribution of {\it measured} magnitudes, $M_0$, is consistent with the 
distribution of {\it true} magnitudes, $M$, in the whole survey.   
This is the underlying principle of the method outlined here.

For  the case {\it (i)} of a local thick shell centered on the MW, a bulk flow $\vB$ yields a galaxy radial peculiar velocity 
$v_{_{\rm B}}= B\cos \theta$ where $\theta$ is the angle between $\vB$ and the line-of-sight to the galaxy. 
This  flow introduces a systematic angular dipolar modulation  
in the distribution of
$M_0-M$ across the sky, which allows a determination of the magnitude  and direction of $\vB$.
For the case {\it (ii)} of a
distant region, only the component of $\vB$ in the direction to the line-of-sight to the subvolume 
is relevant, giving rise to a systematic difference $M_0-M$ 
for all galaxies in the subvolume. In this case only the 
line-of-sight component of $\vB$ can be constrained.

Systematic differences between $M_0$ and $M$ can be appreciated by comparing 
the luminosity distribution of galaxies in the subvolume, i.e. their luminosity function, 
with the luminosity function of the whole survey.
We define the luminosity function, $\Phi(M) $,  expressed in terms of the absolute magnitudes, 
as the number density of galaxies per unit magnitude. 
 We assume that $\Phi(M) $ is well approximated by
a Schechter form  \citep{schechter}
\begin{eqnarray}
\label{eq:shform}
\nonumber
\Phi(M)&=&0.4\ln(10) \Phi^* 10^{0.4(\alpha+1)(M^*-M)}\\
&\times &{\rm exp}\left(-10^{0.4(M^*-M)}\right)\;.
\end{eqnarray}
The normalization  $\Phi^*$ and the shape parameters  $M^*$ and $\alpha$ generally  depend on the 
 galaxies'  type,  { redshift} and band of observation. 
 In terms of the luminosity ($ M=-2.5\log L +const$), this function acquires the simpler form 
\begin{equation}
\label{eq:shformM}
\Phi(L(M))=0.4\ln(10) \Phi^* \left( \frac{L}{L_*}\right)^{1+\alpha}
{\rm exp}\left(-\frac{L}{L_*}\right)\; .
\end{equation}

\subsection{The General Formalism }

%
%
 We quantify the effect of the bulk flow via the probability distribution functions of the magnitudes 
 in the subvolume and the whole survey. 
We write the conditional probability that a galaxy  at redshift $cz$  in the subvolume possesses a {\it measured} 
absolute magnitude $M_0$,
\begin{eqnarray}
\label{eq:pmcz}
&&P(M_0|cz; v_{_{\rm B}})=\int P(M_0|r)P(r|cz)   \dd r  \nonumber \\
                  &&=\int  P(M_0|M)P(M|M_{\rm int})P(M_{\rm int}) P(r|cz)   \nonumber \\
&&\times  \Theta(M_{l}(r)-M) \dd M_{\rm int}\; \dd M\; \dd r\; .
\end{eqnarray}
Here $v_{_{\rm B}}$ is the component of the bulk flow, $\vB$, in the line-of-sight to the galaxy. 
The magnitudes $M_0$ and $M$ are related by Eq.~\ref{eq:shift} and $M_i$
is the { {\it intrinsic}} absolute magnitude which differs from $M$ due to 
photometry errors and small scale peculiar motions not described by the bulk flow. 
The Heaviside step function, $\Theta$,  accounts for the magnitude cut imposed by
the apparent magnitude limit $m_{l}$, i.e. $M_{l}(r)=m_{l}-15 - 5\log r$.
The {{ intrinsic}} magnitude, $M_{\rm int}$, appears only in the underlying luminosity function $P(M_{\rm int}|r)$ and in $ P(M|M_{\rm int}i)$ which accounts for the difference between $M$ and $M_{\rm int}$. 
The probability distribution function  
\begin{equation}
P(M)=\int P(M|M_{\rm int})P(M_{\rm int})\dd M_{\rm int} \; 
\end{equation}
is proportional to the luminosity function of the whole survey,
$\Phi(M)$, and we assume that it can be described well by a Schechter form despite the 
convolution of $P(M_{\rm int})$ with $P(M|M_{\rm int})$.
All other terms in (\ref{eq:pmcz}) are straightforward. We have: 
\begin{equation}
P(M_0|M)=\delta^D(M+\gamma -M_0)\; ,
\end{equation}
where $\delta^D$ is the Dirac $\delta$-function. 
$P(r|cz)$ can be written in a more convenient form using
 Bayes' theorem: 
$P(r|cz)=P(cz|r) P(r)/P(cz)$  where 
\begin{eqnarray}
\label{eq:pr}
P(r)=r^2n(r)& {\rm and} &P(cz)=(cz)^2n(cz) \;,
\end{eqnarray} 
and we assume that the number density of objects is constant along the line of sight,
$n(r)\approx n(cz) \approx const \;,$ and where
\begin{equation}
\label{eq:pczr}
P(cz|r)=\frac{1}{\sqrt{2\pi\sigma_{_{cz}}^2}}{\rm e}^{-\frac{(r+v_{_{\rm B}}-cz)^2}{2\sigma_{_{cz}}^2}}  \; ,
\end{equation}
which assumes that redshifts are normally distributed about the value  $r+v_{_{\rm B}}$.
 The  dispersion is $\sigma_{_{cz}}^2=\sigma_{0}^2+\sigma_{v}^2$
where $\sigma_0$ corresponds to  the $rms$ of errors in  the measured redshifts
and $\sigma_v$ to  small scale motions not described by the bulk flow.
Substituting all this in Eq.~\ref{eq:pmcz} and  integrating
 over $M$  gives
\begin{eqnarray}
\label{eq:pfin}
\nonumber P(M_{0}|cz; v_{\rm B}) &\propto&  
\int_0^\infty r^2 \dd r \Phi\left(M_0-\gamma\right)\\
&\times& \Theta\left(M_{0 l}(cz)-M_0\right) {\rm e}^{-\frac{(r+v_{_{\rm B}}-cz)^2}{2\sigma_{_{cz}}^2}} \; ,
\end{eqnarray}
where the argument of the step function
is now $M_{0l}(cz)-M_0$ where $M_{0l}(cz)=M_{l}(r)+\gamma=m-15-5\log cz$. 

Instead of integrating over  radial coordinates, it is convenient to integrate 
over luminosity  $L(r)\propto 10^{-0.4 M(r)}$. 
Substituting $r= cz (L/L_0)^{1/2}$  and { using} the Schechter 
functional form (\ref{eq:shformM}) for $\Phi(M(L))$ we get
\begin{eqnarray}
\label{eq:pfinL}
\nonumber P(M_{0}|cz; v_{_{\rm B}})& \propto  &
L_0^{-3/2}\int_0^\infty  \dd L  L^{3/2+\alpha}{\rm e}^{-L/L_*}\\
&\times& \Theta\left(\frac{L_0}{L_l(cz)}\right) {\rm e}^{-\frac{(cz(L/L_0)^{1/2}+v_{_{\rm B}}-cz)^2}{2\sigma_{_{cz}}^2}} \; .
\end{eqnarray}
The  conditional probability must integrate to unity,  hence we write the normalized probability for 
the observed $M_0$ as 
\begin{equation}
\label{eq:Pnormz}
P(M_{0}|cz; v_{_{\rm B}})=\frac{0.4 \ln (10)  \; L_0^{1+\alpha}\int_0^\infty  \dd y {\rm e}^{F(y)}  } {\int_{L_l(cz)}^\infty \dd L_0 L_0^{\alpha}\int_0^\infty  \dd y {\rm e}^{F(y)} }\; ,
\end{equation}
where 
\begin{equation}
\label{eq:F}
F(y)\equiv(3+2 \alpha) \ln\;  y -y^2 L_0/L_* -\frac{(y+v_{_{\rm B}}/cz-1)^2}{2(\sigma_{_{cz}}/cz)^2}
\end{equation}
and $y\equiv(L/L_0)^{1/2}$. Note that the expression does not involve the $\Theta$ function. This is because the {\it measured} $M_0$ is derived from observed redshifts and apparent magnitudes and hence 
it is guaranteed that $\Theta\left(M_{ l}(cz)-M_0\right)=1$. 
Equation~\ref{eq:Pnormz} is
the expression for the distribution of the observed magnitudes of galaxies in the subvolume. 
By comparison, the expression for the distribution of the observed magnitudes in the whole survey
is
 \begin{equation}
 P_{\rm survey}(M_0|cz)=\int  P(M_{0}|cz; v_{_{\rm B}}) P(v_{_{\rm B}}) \dd v_{_{\rm B}}\; .
 \end{equation}
 For a Gaussian field $ P(v_{_{\rm B}}) $ is expected to be Gaussian with zero mean { and}
 rms $\sigma_B$. The integration over $v_{_{\rm B}}$ gives 
 a similar expression to (\ref{eq:Pnormz}) but with $v_{_{\rm B}}=0$  and 
 ${\tilde \sigma^2_{_{cz}}}=\sigma^2_{_{cz}}+\sigma^2_B$ instead of $\sigma^2_{_{cz}}$. 
 
The general strategy to estimate a bulk motion should now be clear. It can be described as a two-step porcedure. {\it (i)}
Find the parameters $\alpha$ and $L_*$ of the Schechter function which minimize
{ the quantity}
  $-\sum_{j} \ln P_{\rm survey}((M_{0j}| cz_j)$, where the summation is over all galaxies in the 
  parent survey.   {\it (ii)} Insert these parameters in (\ref{eq:Pnormz}) and search for the
value of $v_{_{\rm B}}$  which minimize
$-\sum_{i}\ln  P(M_{0i}| cz_i; v_{_{\rm B}})$, where now the summation is only over 
galaxies in the subvolume.

The terms $(\sigma_{cz}/cz)^2$ and $({\tilde \sigma{_{cz}}}/cz)^2$  are very small compared to the expected signal $|\vB|/cz$, even in presence of 
photometric errors, so the approximation { is valid for all applications considered in this 
work}.
For $|v_{_{\rm B}}|/cz,\ll 1$ and $\sigma_{_{cz}}/cz \ll 1$ the function  $F(y)$ is very well approximated by a quadratic function in $y$. This fact allows us to apply
the steepest descent method \citep{ARFKEN}
to  evaluate numerator and denominator of (\ref{eq:Pnormz}).

For the numerator we write $F(y)=F_{max} + (1/2)  (y-y_{max})^2/{\cal C_F}^2$, 
where $y_{max}$ is the value where $F(y)$ has a maximum
and ${\cal C_F} \equiv 1/\sqrt{-\dd^2 F/\dd y^2}$ is also evaluated at $y=y_{max}$. 
The steepest descent method yields the following approximation 
\begin{equation}
\int_0^{\infty} \dd y {\rm e}^F=\sqrt{2\pi} {\cal C_F} e^{F_{max}} \; .
\end{equation}

For the denominator we integrate over $ L_0$ and get 
\begin{equation}
\label{eq:deno}
\int_{L_l(cz)}^\infty \dd L_0 L_0^{\alpha}\int_0^\infty  \dd y {\rm e}^{F}
=L_*^{1+\alpha}\int_0^\infty\dd y {\rm e}^{G(y)}
\end{equation}
where 
\begin{equation}
G=\ln \Gamma(1+\alpha, y^2 L_l/L_*) +\ln y -\frac{(y+v_{_{\rm B}}/cz-1)^2}{2(\sigma_{_{cz}}/cz)^2}
\end{equation}
where  $\Gamma(a,x)=\int_x^\infty \dd t {\rm e}^{-t} t^{a-1}  $ is the upper incomplete gamma function. 
The application of steepest descent to the denominator of Eq.~\ref{eq:deno} 
gives the following expression for the conditional probability: 
\begin{equation}
\label{eq:sdp}
P(M_{0}|cz; v_{_{\rm B}})=0.4 \ln (10) \left(\frac{L_0}{L_*}\right)^{1+\alpha}\frac{{\cal C_F}}{{\cal C_G}}{\rm e}^{F_{max}-G_{max}}\; ,
\end{equation}
where $G_{max}$ is the maximum of $G$ and ${\cal C_G}= 1/\sqrt{-\dd^2 G/\dd y^2}$ is evaluated at this maximum.

These approximations are very accurate. We have checked that for the typical values which 
are relevant for this work the accuracy is better than one part in $10^4$.

\subsection{Long distance and small redshift errors approximation}

As a final step we now assume that the amplitude of the bulk flow, small-scale peculiar velocities
and redshift errors are small compared to 
to the typical redshift of the objects in the survey.
In this { hypothesis} we can use (\ref{eq:sdp})  to derive $P(M_0|cz; v_{_{\rm B}})$ to first order in $v_{_{\rm B}}/cz$ and
$\sigma_{_{cz}}/cz$.
The 
 maximum of $F$, to second order 
in $\sigma_{_{cz}}/cz$ and $v_{_{\rm B}}/cz$, is obtained at 
$y_{max}=1-v_{_{\rm B}}/cz +(3+2\alpha-2L_0/L_*)(\sigma_{_{cz}}/cz)^2$.
Substituting this expression into (\ref{eq:F}) we find, 
to first order terms in $v/cz$, that
\begin{equation}
{\rm e}^{F_{max}}=\left(1-(3+2\alpha)\frac{v_{_{\rm B}}}{cz}\right){\rm e}^ {-(1-2v_{_{\rm B}}/cz)L_0/L_*}\; ,
\end{equation}
and similarly 
\begin{equation}
{\rm e}^{G_{max}}=\left(1-\frac{v_{_{\rm B}}}{cz}\right)\Gamma\left(1+\alpha, (1-2v_{_{\rm B}}/cz)L_l/L_*\right)\; .
\end{equation}
Further, we get 
$ {\cal C_G}^2/{\cal C_F}^2=1$.
Inserting all these expression into  (\ref{eq:sdp}) gives
\begin{equation}
\label{eq:papp}
P(M_0|cz; v_{_{\rm B}})=\frac{0.4 \ln (10) \left(\frac{{\tilde L}_0}{L_*}\right)^{1+\alpha}{\rm e}^ {-{\tilde L}_0/L_*}}{\Gamma\left(1+\alpha,{\tilde L}_l/L_*\right)}
\end{equation}
where ${\tilde L}_0=(1-2v_{_{\rm B}}/cz)L_0$ and ${\tilde L}_l=(1-2v_{_{\rm B}}/cz)L_l$.  
In practice, the effect of a bulk flow $v_{\rm B}$ is that of shifting the estimated luminosity of
a galaxy $L_0$ by a factor which is proportional to the amplitude of the flow.
If applied to a distant subvolume moving at velocity $v_{\rm B}$, the net effect would be that 
of shifting the typical luminosity $L_*$ of the {\it measured} luminosity function, with no 
impact on the slope $\alpha$ and normalization $\Phi_*$.
This result is hardly surprising, but it is reassuring to see it  obtained in the 
limit of small velocity $|v/cz|$ and errors  $\sigma_{_{cz}}/cz$.
Second order corrections to this expression involve
terms of $O((\sigma_{_{cz}}/cz)^2)$ and $O((v_{_{\rm B}}/cz))^2$, which are typically very small 
when one considers cosmological volumes.

%
%

\section{Application to 2MRS}

In this section we apply the method outlined above to the 2MRS.
The goal is to estimate the bulk flow of a thick spherical shells centered on the MW
at $ z \sim0 $.
2MRS is an all-sky redshift catalog of about 23,200 galaxies, which is complete down to the 
magnitude  $K=11.25$. 
Details about the catalog, including the precise completeness, sky coverage and selection effects can be found in
 \citep{fhuch}. The preparation of the catalog for the purpose of the application of the method is 
is done similarly to \cite{ND11}.

\begin{figure*} 
\centering
\epsscale{1}
\plotone{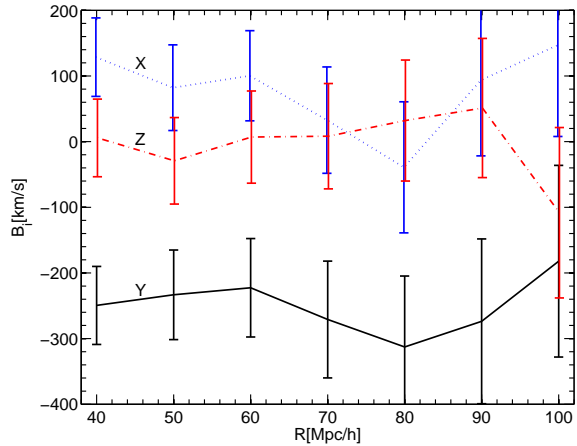}
\caption{The bulk flow of $40\hmpc$-thick shells centered on the Milky Way. The $X$, $Y$ and $Z$ components of the 
bulk flow are plotted against the mean radius of the shell, as indicated in the figure.  }
\label{fig:Bdiff}
\end{figure*}

The description of the method  presented in the previous Section
assumes that the luminosities of galaxies in the sample are drawn from 
a single luminosity function. 
However, the method could easily be extended to 
account for the different  luminosity functions of different galaxy types
(e.g. red vs. blue, spirals vs. ellipticals) and environments (high density vs. low density regions).
For this reason we have divided galaxies in two samples: spirals and ellipticals, and measured their
luminosity functions separately. { We find $(\alpha, M_*)=(-0.803,-23.53)$ for early-type and 
$(\alpha, M_*)=(-0.888,-23.12)$, for late-type galaxies in agreement with the previous, independent estimates 
by \cite{w09}. We made no attempt to determine $\Phi^*$ which is not relevant for our goal.} 
These are the reference values for the luminosity function of the whole survey.
To determine the bulk flow we consider a spherical subvolume of radius
 $ cz=10000\kms$ that contains 16460 galaxies brighter than $K=11.25$, of which
10366 and 6094 are late and early types, 
respectively. We divided this sample
into spherical shells each one $4000\kms$ thick. 
To estimate the three  Cartesian components of the  bulk flow, $\vB$, in each shell, 
we have minimized  the probability function $-\sum_i \ln P(M_{0i}|cz_i; v_{_{\rm B}})$  with respect to $\vB$, where 
the summation is over all galaxies in the shell.
The expression for  (\ref{eq:papp}) is used for $P$ where different  values of $\alpha$ and $L_*$ 
are used for late and early type galaxies. 
The results are shown in 
Fig.~\ref{fig:Bdiff}.
{ The points represent the Cartesian components of the bulk flow, indicated on the plot, 
estimated in correspondence of the mean radius of each shell.} 
The  $1\sigma$ errorbars were obtained from 200 
mock samples mimicking the 2MRS subvolume. 
Each sample contains a number of points similar to that of the 2MRS galaxies. 
{ The points are randomly distributed  within a sphere of $200\hmpc$} 
and are assigned absolute magnitudes according to the overall reference luminosity function of all 2MRS galaxies.
{ Mock galaxies were assigned random peculiar velocities 
sampled from a Gaussian distribution with zero mean and  a $300\kms$ scatter 
to mimic  the combined effect of small scale velocity dispersion
and errors in the measured redshift. No bulk flow was assigned to the mock samples.}
Our minimization procedure was then applied to the 200 mock samples to derive 
the intrinsic uncertainties in the estimate of the bulk flow. These uncertainties are plotted as 
$1\sigma$  errorbars. 

Despite the large errors on the bulk from K=11.25 2MRS,  the results  are very encouraging. 
  They provide a strong incentive for 
a more thorough study using the deeper K=12.2 2MRS which will 
triple  number of galaxies within $cz= 10000\kms$.
The results can be compared with 
bulk flow  of \cite{ND11} (see their figure 5) obtained using the Tully-Fisher SFI++ data.
A full comparison between the photometric and the SFI++ results 
is complicated and should take into account the complex covariance of errors
in both measurements. The effort is redundant at this stage due to the large errors 
on the current 2MRS photometric bulk flow. 
Therefore,   we only make a comparison between the two bulk flows 
at $R\approx60\hmpc$. At this intermediate radius the photometric bulk 
$\vB_{\rm phot}\approx(100\pm 90, -240\pm 90, 0\pm 90)\kms$ and the SFI++
 $\vB_{\rm SFI++}\approx (15\pm 20, -280\pm 30, 120\pm 20 )$  (see top panel of figure 5 in \cite{ND11}) are 
fully consistent. 
We note that $\vB_{\rm phot}$ refers to  shells of thickness $40\hmpc$, while $\vB_{\rm SFI++}$
to the full sphere within $R$. But this hardly matters since $\vB_{\rm phot}$ versus distance is nearly flat
within the errors. 
The reason that $\vB_{\rm phot} $ is computed for  shells rather than the whole spheres is 
that the bulk is heavily weighted  by the lower redshifts galaxies due to the rapid decrease 
in the number of 2MRS galaxies at larger redshfits.
Computing the photometric bulk in spherical shells allows a clearer inspection of the  constraints we get at larger 
redshifts. 
In contrast, the SFI++ flow within a sphere of radius $R$ is, however, mainly dominated 
by galaxies nearer to $R$.

The method outlined above is not the only one to estimate the bulk velocity of a spherical volume 
centered on the MW. 
We have implemented an alternative procedure based on the fact that
a  bulk flow, $\vB$, 
introduces  a dipole-like angular modulation in $M_0$ across a spherical shell centered on the observer.
For $|\vB|/cz\ll 1$, 
$M_0=M-5\log e \; \vB \cdot \hat {\vr}/cz =-2.1715 \vB \cdot \hat {\vr}/cz$, 
where $\hat {\vr}  $ is the usual unit vector in the line-of-sight to the galaxy.
The bulk flow, $\vB$,  can then be found by minimizing
$
\chi^2= \sum( M_{0i}+ 2.1715 \vB \cdot \hat {\vr}_i)^2\; .
$
We have checked that this alternative 
strategy  gives results that are noisier than but consistent with the luminosity-based method
which uses the shape of the luminosity function, in addition to the measured magnitudes and redshifts.

\section{Prospects for future surveys} 

In this section we assess the potential of our method applied to future datasets.
The main goal of most of the next-generation large redshift survey 
is to constrain Dark Energy properties. For this reason, they are 
designed to optimize observations at $z\sim 1$, i.e the epoch at which Dark
Energy started dominating the energy budget of the Universe.
However, some of these projects will also provide redshifts, measured either
spectroscopically or through multi-band photometry, for a large number of 
relatively local galaxies (i.e. $z\le 0.5$), where  
the bulk flow can be reliably estimated by our photometry-based method. 
Among the planned redshift surveys that meet these constraints are
the ground-based LSST project \citep{2009arXiv0912.0201L} and the 
EUCLID satellite mission  \citep{euclid09}. The two surveys enjoy the 
same large sky coverage ($\sim 20000$ deg$^2$) and are expected to 
deliver photometric redshifts for a large (2-4 billion) number of galaxies 
in the redshift range $z=[0,3]$. In addition, the EUCLID project is also expected to
measure spectroscopic redshifts for about 70 million galaxies. 
Given the present uncertainties in the survey design we do not attempt here 
to provide an accurate forecast for the possibility of measuring our bulk flows 
from these surveys. Instead, we will
allow for generous variations in the 
survey design parameters and provide an order of magnitude estimate
for our capability to constraint the bulk flow at $z\le 0.5$.
For this purpose we will use the EUCLID
survey as a baseline and allow 
for significant variations in the number of the objects in the survey
which should include the case of the LSST survey.

To model the characteristic of both redshift surveys, the spectroscopic 
and the photometric ones, we need to specify the luminosity function of the galaxies,
their redshift distribution and the errors on the photometry and the redshift.

For the luminosity function one we consider a Schechter 
form in the H$_{\rm AB}$-band with values of $\alpha$ and  $M_*$ derived 
from \cite{2010MNRAS.401.1166C} assuming the H-K color correction 
proposed by \cite{1999ApJ...511...16J}.
The luminosity function evolves with redshift. Here we are interested in 
two epochs:  $z=0.1$ and $=0.5$, for which we find 
$(\alpha, M_*)=(-1.07,-22.15)$ and 
$(\alpha, M_*)=(-1.07,-22.21)$,respectively.
The spectroscopic catalog is expected to contain objects
with apparent magnitude brighter tha  H$_{\rm AB}=19.5$ while the photometric catalog
will sample galaxies as faint as H$_{\rm AB}=24$. We assume a photometric error 
$\sigma_m=0.2$ which  represents the expected error for faint objects
and thus overestimate the uncertainties for brighter objects.
We note that the expected number of objects in the EUCLID photo-z catalog
($\sim 10^5$ gals. deg$^{-2}$ with H$_{\rm AB}<24$) is remarkably similar to
that  in the LSST  one ($\sim 2 \times  10^5$ gals. deg$^{-2}$ with i$_{\rm AB}<25.3$).
A similarity that also applies to the expected galaxy redshift distribution and that further justifies our choice
of not considering the two surveys separately.

The galaxy redshift distribution $dN/dz$ for the EUCLID spectroscopic catalog has been 
modeled by \cite{2010MNRAS.402.1330G}. In the current design, the 
survey will target ${\rm H} \alpha$
 galaxies above  a limiting flux  $f_{{\rm H} \alpha} = 3\times10^{-16}$  $
\textrm{erg } \textrm{cm}^{-2}  \textrm{s}^{-1}$. The $dN/dz$ of these objects is 
listed in Table 2, column 3 of that paper. The numbers must be further divided by 
a factor $\sim 4$ to account for the efficiency
in the determining the redshift from the ${\rm H} \alpha$ line 
(Guzzo, private communication). This survey will measure the spectroscopic
redshift of  objects with $z>0.45$ with an expected accuracy $\sigma_z=0.001$.
In the redshift bin $\Delta z=0.1$ centered at $z=0.5$, the expected number of 
galaxies is $N\sim 4 \times 10^6$. To simplify our calculation we adopt a more 
conservative $N=10^6$ and also explore the pessimistic case 
of measuring only  $N=10^5$ redshifts.

The expected number of galaxies in the photo-z catalog
can be derived by the observed galaxy $dN/dz$ the GOODS and UKIDDS-UDS fields 
(Zamorani, private communication). This catalog will include low-redshift objects 
and redshift will be measured with an accuracy $\sigma_z/(1+z)=0.05-0.03$ (expected-goal). Here we adopt 
the first, more conservative estimate.
At $z=0.5$ the number of objects in this catalog is $\sim$ 10 times larger than in the spectroscopic one, 
i.e. the same shell at $z=0.5$ considered above will have $N=10^7$ objects with measured photo-z.
As before, we will  also consider two pessimistic cases in which the number of measured 
redshifts is   $N=10^6$ and  $N=10^5$.
In the following we will be also interested in measuring the bulk flow in the 
nearby universe, at $z=0.13$. At this distance EUCLID is expected  to measure 
$N=10^6$ photo-z in a sphere of radius $100$ $h^{-1}$ Mpc. We consider this 
as a reference number density at  $z\sim0.1$ but also consider a pessimistic and an 
optimistic cases in which $N=10^5$ and $N=10^7$, respectively.

To investigate whether our method can be applied successfully 
to these next-generation surveys  we 
consider two possible estimates of the bulk flow: at $z=0.13$
and at $z=0.5$. As a first example we consider a bulk flow $v_{_{\rm B}}=300\kms$ 
in a sphere of radius $100$ $h^{-1}$ Mpc  at  $z=0.13$, 
similar in size to the one we have considered  in the 2MRS sample.
As a second case we consider a  large
bulk flow $v_{\rm B}=1000\kms$ 
of a  shell at  $z=0.5$ with $\Delta z=0.1$, reminiscent of the recent claimed detection by 
\cite{2010ApJ...712L..81K} and potentially conflicting with the standard cosmological framework.
These two bulk flows introduce the same systematic shift in the apparent magnitude of the 
objects: $\delta M\approx 0.018$. Our task is to assess if  the next generation surveys 
described above will be able to detect such shift. 
To summarize, we will explore the following scenarios:
\begin{itemize}

\item The case of a spherical region of radius  $100$ $h^{-1}$ Mpc moving at $v_{_{\rm B}}=300\kms$ 
at $z=0.13$, containing $N=10^6$ objects with measured photo-z. In addition to this reference case
we also consider an optimistic and pessimistic scenarios in which the number of measured photometric 
redshifts is $N=10^7$ and $N=10^5$, respectively. The results relative to this scenario are listed in 
the first 12 rows of Table~\ref{tab:error}.

\item The case of a shell at  $z=0.5$ centered on the MW, with $\Delta z=0.1$ and moving $v_{\rm B}=1000\kms$,
sampled by $N=10^7$ objects with measured photo-z (reference case). We also consider the pessimistic 
cases of $N=10^6$ and $N=10^5$ objects. The results relative to this scenario are listed in 
rows 13-24 of Table~\ref{tab:error}.

\item The case of the same shell at  $z=0.5$ moving at the same speed but sampled by $N=10^6$ with redshift measured spectroscopically.
We also consider a more pessimistic case of  $N=10^5$ objects. The results are listed in rows 25-26  of Table~\ref{tab:error}.
\end{itemize}

The general strategy to assess our ability in measuring the bulk flow is straightforward.
First, we consider a very large set of objects representing the whole survey to which we assign absolute magnitudes according to the reference 
luminosity function and determine the $\alpha$ and $M_*$ by minimizing  $- \sum_j P_{\rm survey}(M_{0j}|cz) $.
Then, for each of the three scenarios itemized above,  we repeat the exercise considering 70 subsamples, i.e. we determine 
a  set of ($\alpha$,$M_*$) values by minimizing  $-\sum_jP(M_{0j}|cz; v_{\rm B}=0) $. Objects in the subsamples  are assigned a 
magnitude error by adding a Gaussian scatter  of zero mean and width $\sigma_M$ and no bulk flow.
The condition for a significant bulk flow detection is that the rms scatter in the value of $M_*$ measured from the 70 mock samples 
is larger than the expected magnitude shift $\delta M\approx 0.018$

The results of this test are summarized  in Table~\ref{tab:error}. They refer to the three scenarios outlined above. 
To assess the impact of photometric errors we have explored the case of $\sigma_M=0$ (no photometric errors) and that
of fixing $\alpha$  at its  mean value as derived from the mean of subsamples.
In this Table,  $\mudm$ is the difference between the mean of the recovered  $M_*$ from the 70  subsamples and the true value
and  its scatter $\sigdm$  . The quantities  $\mual$ and $\sigal$ are defined 
similarly but for $\alpha$.

For $N=10^7$, we get very small  $\sigdm$ and $\sigal$. This means that one can treat the 
corresponding values  as those that would be  obtained from the full survey (which is expected to contain a 
significantly large number of galaxies than $10^7$). Since the presence of a bulk flow is not expected to affect 
$\alpha$ but only  $M_*$, this result suggests that one can safely fix $\alpha$ to the value measured from the whole survey, i.e. 
that the case which is directly relevant for our method is the one tagged 'fixed $\alpha$ in Table~\ref{tab:error}. We also show 
the case of free  $\alpha$ for the sake of completeness.
The results also show that when $\sigma_M=0$ the values of 
$\mudm$ and $\mual$ vanish. This is a known effect and reflects the fact that
random photometric errors preferentially scatter galaxies to the brighter end of the 
luminosity function. 
However, the bias would affect the estimate of the {\it true} luminosity
function from the whole  survey and hence it is irrelevant to our method 
which only uses the values of $\alpha$ and $M_*$ measured from the 
full survey in order to get the bulk flow by minimization of   $-\sum_jP(M_{0j}|cz; v_{\rm B}) $.

The results show that the bulk flows that we are considering here and
which would introduce a systematic magnitude shift 
$\delta M =0.018 $ would be detected at high significance 
in all three reference cases considered
(boldface quantities in Table~\ref{tab:error}).
In particular, we see that the $300\kms$ bulk flow at $z=0.13$ would be 
detected at $\sim 5 \sigma$ significance by measuring the photometric redshift of
$10^6$ objects in a a sphere of  $100$ $h^{-1}$ Mpc.
At $z=0.5$, $10^7$ photometric redshifts would allow us to detect 
a bulk flow of $1000\kms$ at $\sim 20 \sigma$, decreasing to 
$\sim 7 \, {\rm and} \, 3 \sigma$, respectively, for  $10^6$ and $10^5$
photometric redshifts.
The statistical significance of the bulk flow detection
can be further increased by reducing redshift errors, i.e. 
by using spectroscopic rather than photometric redshifts. 
The result is seen by comparing 
the two cases of a spectroscopic and a photo-z with the same value of $N$ and $\sigma_M$,
(e.g. row 25 vs. row 20 in Table~\ref{tab:error}). The value of $\sigma_M$ is reduced by a factor 
of $\sim 2$. 
The reason of this improvement  is that galaxies in the spectroscopic 
catalog sample the brighter end of the luminosity function which is 
sensitive to the value of  $M_*$. The result is that,  for the same value of $N$,
the galaxies in the spectroscopic sample are able  
to constrain $M_*$ considerably  better than galaxies in the photometric catalog.
The same reasoning explains why  
$\sigdm$ in row 16, which refers to a sample of distant, bright galaxies
at $z=0.5$ is smaller than  the value in row 4 which refers to a sample of 
nearer and fainter galaxies.

\begin{table*}
\begin{center}
\label{tab:error}
\caption{\label{tab:error}{\it Expected error on $ M_*$ from a survey similar to  EUCLID}.   
 Col.2: redshift. Col 3: number of galaxies in the subvolume. 
 Col 4: assumed error on the magnitude. Col 5: deviation from 
 the actual $M_*$ written as the systematic bias $\pm$ the $1\sigma$ error.
 For ``fixed $\alpha$", the value of $\alpha$ was fixed from the whole survey.
 Rows 1--24 and 25--26  refer, respectively,  to the photometric and spectroscopic 
 redshifts catalogs.   The differences results between these catalogue for the same $N$ are due to the different apparent magnitude limits:   H$_{\rm AB}=24$ for the photometric and 
  H$_{\rm AB}=19.5$ for the spectroscopic catalog. Most relevant cases for EUCLID are emphasized in boldface. }
\bigskip
\begin{tabular}{|c|c|c|c|c|c|c|}
\hline \hline
 row \# & $z$  &  $N$ & $\sigma_M$&  $\mu_{_{\delta\! M_*}}\pm \sigma_{_{\delta\! M_*}}$  & 
$\mu_{_{\delta\! \alpha}}\pm \sigma_{_{\delta\! \alpha}}$ \\ 
\hline 
 1& 0.13   & $10^7$ & $0$   & $0\pm 0.0013$  &  $0\pm 0.0002$\\  
 2& 0.13   & $10^7$ & $0$   & $0\pm 0.0009$ & fixed $\alpha$ \\ 
 3& 0.13   & $10^7$ & $0.2$ & $-0.035 \pm 0.0016$ & $-0.0028 \pm 0.0002 $\\  
 4& 0.13   & $10^7$ & $0.2$ & $-0.035\pm 0.0013$ & fixed $\alpha$ \\ 
\hline
 5& 0.13   & $10^6$ & $0$   & $0\pm 0.0041$  &  $0\pm 0.0007$\\  
 6& 0.13   & $10^6$ & $0$   & $0\pm 0.0031$ & fixed $\alpha$ \\ 
 7& 0.13   & $10^6$ & $0.2$ & $-0.035 \pm 0.0056$ & $-0.0028 \pm 0.0007$ \\  
 {\bf  8}& \bf 0.13   & $\bf 10^6$ & $\bf 0.2$ & $\bf -0.035\pm 0.0037$ &\bf fixed $\alpha$ \\ 
\hline
 9&0.13   & $10^5$ & $0$   & $0\pm 0.015$  & $ 0\pm 0.0025$\\  
10&0.13   & $10^5$ & $0$   & $0\pm 0.0011$ & fixed $\alpha$ \\ 
11&0.13   & $10^5$ & $0.2$ & $-0.035 \pm 0.0158$ & $-0.0028 \pm 0.0021 $\\  
12&0.13   & $10^5$ & $0.2$ & $-0.035\pm 0.0114$ & fixed $\alpha$ \\ 
\hline
13&0.5  & $10^7$ & $0$   & $0\pm 0.0014$      &  $0\pm 0.0005$\\  
14&0.5  & $10^7$ & $0$   & $0\pm 0.00075$     & fixed $\alpha$ \\ 
15&0.5  & $10^7$ & $0.2$ & $-0.05 \pm 0.0015$ & $-0.01\pm 0.0005$\\ 
{\bf  16}&\bf 0.5  & $\bf 10^7$ & $\bf 0.2$ & $\bf -0.05\pm 0.0008$  & \bf fixed $\alpha$ \\
\hline
17&0.5  & $10^6$ & $0$ & $0\pm 0.0048$ &  $0\pm 0.0018$\\  
18&0.5  & $10^6$ & $0$ & $0\pm 0.0024$ & fixed $\alpha$ \\ 
19&0.5   & $10^6$ & $0.2$ & $-0.05 \pm 0.0047$ & $ -0.01\pm 0.0016$\\  
20&0.5    & $10^6$ & $0.2$ & $-0.05\pm 0.0026$ & fixed $\alpha$ \\ 

\hline
21&0.5  & $10^5$ & $0$ & $0\pm 0.013$ & $0\pm 0.0049$\\  
22&0.5  & $10^5$ & $0$ & $0\pm 0.008$ & fixed $\alpha$ \\ 
23&0.5   & $10^5$ & $0.2$ & $-0.05 \pm 0.0136$ & $-0.01\pm 0.005 $\\  
24&0.5    & $10^5$ & $0.2$ & $-0.05\pm 0.0065$ & fixed $\alpha$ \\ 
\hline\hline
{\bf  25}&\bf 0.5  & $\bf 10^6$ & $\bf 0.2$ & $\bf -0.08\pm 0.0015$ & \bf fixed $\alpha$ \\  
26&0.5  & $10^5$ & $0.2$ & $-0.08\pm 0.0044$ & fixed $\alpha$\\ 

\hline
\end{tabular}
\end{center}
\end{table*}




\section{Discussion}

Over the years cosmological  bulk flows has been estimated from 
the measured peculiar velocities of a large variety of objects ranging from galaxies
\citep{1998ApJ...505L..91G, 1998AJ....116.2632G, 1999ApJ...522....1D, 2000ApJ...544..636C, 2000ApJ...537L..81D,
2007MNRAS.375..691S}
clusters of galaxies \citep{1994ApJ...425..418L, 1996ApJ...461L..17B, 2004MNRAS.352...61H}
and SNIa  \citep{1995ApJ...445L..91R}. Conflicting results triggered by the use of error-prone distance indicators 
have fueled a long lasting controversy on the amplitude and  convergence of the bulk flow that is still on.
For example, using the SFI++ galaxy catalog \citep{2009MNRAS.392..743W},
claimed the detection of a bulk flow of $407\pm81$ km s$^{-1}$ within $R=50$ $h^{-1}$Mpc,
inconsistent with expectation from the $\Lambda$CDM model. This result
has been challenged by the re-analysis of the same data by \cite{2011arXiv1101.1650N} who found a 
bulk flow amplitude consistent  with  $\Lambda$CDM expectations.
Several bulk flow estimates have been recently performed from the dipole-like anisotropy induced
by the kinetic Sunyaev-Zel'dovich decrement in the WMAP temperature map, measured at the 
position of X-ray galaxy clusters. When interpreted  as a coherent motion, this signal would indicate 
a  gigantic bulk flow of $1028\pm 265$ km s$^{-1}$ within $R=528$ $h^{-1}$Mpc, clearly inconsistent
with the standard picture of gravitational instability. However, a more recent and independent analysis of 
 WMAP data \citep{2010arXiv1011.2781O} did not confirm this result.
Finally, systematic anisotropies in the observed clustering of objects induced by peculiar velocities, the so called
redshift space distortions, are a popular and reliable way to estimate  the growth rate of density fluctuations
and to constrain the amplitude of large-scale coherent motions
\citep{2001Natur.410..169P, 2008Natur.451..541G,
2010MNRAS.406..803B}. However, their interpretation in terms 
of bulk velocity is not straightforward.

Here we have presented and implemented a method to estimate cosmological bulk flows 
that only depends on photometry and redshift measurements and not 
on distance indicators.
The  main requirements for a successful application of this statistical method 
is dense sampling over large regions of the universe.
A large number of object is mandatory since, for a given bulk flow amplitude, 
errors are dominated by Poisson noise, as clearly indicated by 
the fact that the error on $M_*$,  $\sigdm$, listed in 
Table~\ref{tab:error}  scales as $\sqrt{N}$.
We find that the accuracy in the measurement of galaxy magnitude and redshifts 
are less crucial in the bulk flow estimate.
This fact indicates that photometric-redshift catalogs are better suited
for this analysis than the spectroscopic ones which typically include significantly less galaxies.
Of course,  this assumes that 
systematic errors and the fraction of $3\sigma$ outliers
(which can be significant at $z<0.5$  \cite{2009arXiv0912.0201L})
in the photo-z measurement can be conveniently reduced. Dedicated
analysis that accurately mimic the expected performance of the
surveys will be required to address this issue.

Large volumes and dense sampling are requirements that will be met 
by next-generation redshift surveys like EUCLID and LSST. Therefore, we have 
carried out a  feasibility study  for such surveys. Taking into account that 
the sensitivity of our method decreases with the redshift we have only considered
galaxies with $z<0.5$. We  focused on two different measurements: a 
bulk flow of $\sim 300\kms$  at $z\sim 0.15$ on scales $\sim 100 \hmpc$
and a much larger bulk flow of $ \sim 1000\kms $ at $z\sim 0.5$ over a 
$\Delta z=0.1$ shell centered on the MW.
 Both flows would introduce the same 
shift in the measured magnitude of the galaxies.
The results, summarized in Table~\ref{tab:error} are very encouraging.
They suggests, for example, that one could obtain independent estimates
of cosmological  bulk flows (with typical amplitude of $\sim 300\kms$) 
over the $\sim 50$  2MRS-sized samples within $z=0.2$
It is worth stressing that photo-z are accurate enough  for our purposes.
A comparison between  rows 8 and 25 in the Table shows that  $10^6$ galaxies (with photo-z) 
the error in determining $M_*$ is actually smaller at $z\sim 0.5$ than $z\sim 0.13$. 
This implies that a bulk flow of $200-300\kms$ (which amounts to $\delta M\sim  0.0045$ at $z\sim 0.5$ and $\delta _M\sim 0.012$ at $z\sim 0.13$)  is constrained at the  $3\sigma $ level at both redshifts.

As a preliminary application of the method, we have resorted to the  2MRS all sky survey 
which is limited to $K=11.25$. 
The results are consistent with the bulk flow derived by \cite{ND11} using the SFI++ Tully-Fisher 
catalog of distance indicators. The 2MRS photometric bulk flow derived here is 
noisy due to the relatively small number of galaxies. Better results will be obtained 
using the deeper  $K=12.2 $  2MRS. This already planned extension of the original 
survey will increase the number of observed redshifts by a factor of 3 within a distance of $100\hmpc$.
Since Poisson noise dominates the error budget, we expect a factor of $\sqrt{3}$ reduction 
in the size of the errorbars.

There are strong indications that the luminosity function, $\Phi$,  depends on  the 
large scale environment of galaxies \citep[e.g.][]{balogh01,hutsi,crotL05,park07,moyang04}. These 
studies explore the dependence on the density of galaxies on scales of  a few Mpcs.
As outlined above, we are interested in significantly larger scales of  $\gtsim 100\hmpc$.
When averaged over these scales, the rms density fluctuations is 
 less than 0.07  (for standard $\Lambda$CDM model with $\sigma_8=0.8$).
So far no study has addressed environmental dependences of galaxy luminosities over 
such large scales. 
If the overdensity of the large scale environment is the main relevant  parameter, then an extrapolation of the observed dependence  on a few Mpc scales to 100 $\hmpc$ yields an 
exceedingly  small effect. 
Moreover,   in next generation surveys we could easily account for such 
 effects thanks to
our ability to measure the bulk flow over independent volumes that could be 
classified according to their average density.
To illustrate the robustness of our 2MRS result over environmental effects, we note that 
a 300$\kms$ bulk flow at $60\hmpc$ gives a magnitude 
difference of $\delta M= 0.2$ between the apex and the anti-apex of the bulk motion.
In $\Lambda$CDM, the rms density in spheres of $60\hmpc$ is 0.13. Assuming again that the overdensity of the
environment is the only relevant parameter, we find that  $\delta M$ 
should dominate the environmental dependence  extrapolated from smaller scales. 
A further evidence that environmental effects did not affect our bulk flow estimates 
from 2MRS is that the method here gives results that are  
 consistent with those
\cite{ND11} who derive the bulk flow based on  the Tully-Fisher distant indicators,
rather than photometry.

\section{Acknowledgments}

 This work was supported by THE ISRAEL SCIENCE FOUNDATION (grant No.203/09), the German-Israeli Foundation for 
Research and Development,  the Asher Space Research
Institute and  by the  WINNIPEG  RESEARCH FUND.
MD acknowledges the support provided by the NSF grant  AST-0807630. 
EB  acknowledges the support provided by the Agenzia Spaziale Italiana (ASI, contract N.I/058/08/0)
EB thanks the Physics Department and the Asher Space Science Institute-Technion
for the kind hospitality.

\bibliography{photo_APJ}

\end{document}